\documentclass[aps,prl,twocolumn,showpacs,preprintnumbers,amsmath,amssymb]{revtex4}

\usepackage{amssymb,amsmath}
\usepackage[dvips]{graphicx}

\usepackage{dcolumn,epsfig}

\newcommand{\beq}{\begin{equation}}
\newcommand{\eeq}{\end{equation}}
\newcommand{\bes}{\begin{subequations}}
\newcommand{\ees}{\end{subequations}}
\newcommand{\bea}{\begin{eqnarray}}
\newcommand{\eea}{\end{eqnarray}}
\newcommand{\ba}{\begin{array}}
\newcommand{\ea}{\end{array}}
\newcommand{\beqn}{\begin{eqnarray*}}
\newcommand{\eeqn}{\end{eqnarray*}}

\newcommand{\f}[2]{\frac{#1}{#2}}
\newcommand{\g}{\gamma}

\newcommand{\la}{\langle}
\newcommand{\ra}{\rangle}

\def\nn{\nonumber}

\newlength{\sizeonefig}
\newlength{\sizetwofig}
\begin{document}

\title{ Langevin dynamics in crossed magnetic and electric fields: Hall and diamagnetic fluctuations}

\author{Dibyendu Roy and N. Kumar} 
\email{dibyendu@rri.res.in, nkumar@rri.res.in}
\affiliation{Raman Research Institute, Bangalore 560080, India.} 

\begin{abstract}
Based on the classical Langevin equation, we have  re-visited the problem of orbital  motion of a charged particle in two dimensions for a normal  magnetic  field crossed with or without an in-plane electric bias. We are led to two interesting fluctuation effects: First, we obtain not only a longitudinal  ``work-fluctuation'' relation as expected for a barotropic type system,  but also  a  transverse work-fluctuation relation perpendicular to  the electric bias. This  ``Hall fluctuation''  involves the product of  the electric  and the magnetic  fields. And second, for the case of  harmonic confinement without bias, the calculated probability density for  the orbital  magnetic moment gives non-zero even moments, not derivable as field derivatives of the classical free energy. 
\end{abstract}
\date{\today}
\maketitle
Consider the classical motion of a charged particle in a plane under the influence of a normal (out-of-plane) magnetic field crossed in general with a parallel ``in-plane" electric field, in the presence of a dissipative coupling to the environment (bath). This 2D motion has two notable general features -- first, the cyclotron circular orbits about the  magnetic field  under
the Lorentz force that classically does no work, and second, there is the well known transverse drift identified as
the Hall effect which is a dissipative transport (non-equilibrium) phenomenon. The orbital motion holds surprises--the best known of these being the absence  of classical orbital diamagnetism in equilibrium, as embodied in
 the classic theorem due to  Bohr and  van Leeuwen, 
and, indeed, regarded by some to be a surprise of theoretical
 physics \cite{peierls}. A notable point  here
 is the subtle role played by the spatial  confinement (the
 boundary) that interrupts the otherwise complete circular (diamagnetic) orbits leading to a skipping cuspidal motion at the boundary which is retrograde (paramagnetic) and exactly cancels the bulk diamagnetic contribution \cite{darwin30, amj81, sdg97}. Diamagnetism, or rather its absence classically,  \cite{van32} is,
 of course, a well known equilibrium  phenomenon. We, however, find  interesting  fluctuation  effects here. Thus, while the orbital diamagnetic moment is zero in the mean, its higher (even) moments do exist under confinement, but can not be obtained as field derivatives of the classical free energy. The other surprising effect manifests in  the non-equilibrium   steady state as a transverse work-fluctuation relation in addition to the well known longitudinal work-fluctuation theoretic results of much current  interest \cite{evans02, Jarzynski97, Crooks99, Narayan04}. In the light of these, we  re-visit the classical   problem of orbital motion of a charged particle in two dimensions in the  presence of  a normal  magnetic   field. First, we will consider the work-fluctuations in the presence of a parallel  (in-plane) electric bias. Keeping in mind the non-equilibrium steady-state situation of interest here, we follow the Einsteinian statistical approach based on the Langevin
  equation whose solution in the long-time limit  gives the non-equilibrium
 steady-state  results for the system driven externally (e.g., biased electrically). The main results   are the following:  For  non-zero electric bias, a ``work fluctuation" expression obtains as, of course, expected for a barotropic-type
system. In addition, however, a transverse work-fluctuation is also obtained    
 perpendicular to  the electric bias -- a ``Hall fluctuation" theorem. This ``Hall fluctuation"    involves the product of  the electric  and the magnetic
 fields. Inasmuch as classically the magnetic field does no work on a moving charge, this cross effect (Hall fluctuation) seems to be related to the time-reversal symmetry breaking effect of the externally applied magnetic field. This is quite distinct from the time-reversal asymmetry (irreversibility) arising out of dissipation. It is apt to point out at this stage  that  
the physical scheme we have in mind here is experimentally an obvious   
 variation on the well known  Haynes-Shockley \cite{haynes51} set-up for studying the drift concomitant with the spread (diffusion) of minority  charge
 carriers  photo-injected in an electrically biased semiconducting sample.
 Indeed, the fact that the minority carriers can disappear through              
the electron-hole recombination  process, provides for  a study of finite life-time effects in our non-equilibrium steady
state \cite{amj81}. This should be an interesting way of probing  the effect of confinement, or the boundary, that may not be effective when the carrier life-time is very short. 

$Fluctuation~ theorem$ -- $The~classical~ Hall~ bar:$ 
Consider  a classical 2D system of noninteracting electrons (charge $-e)$ under the externally applied crossed electric and magnetic fields, in the presence of a dissipative environment at temperature $T$ (the bath) in the high friction limit.  Let the electric field $(E)$ be 
in the $x-$direction and the magnetic field $(B)$ in the $z-$direction. We can now write down the Langevin equations of motion for the electron in the overdamped limit, i.e., ignoring the inertial effects \cite{Coffey04} as: 
\bea
\g \dot{x}&=& -eE-\f{eB}{c} \dot{y} +\eta_x(t) \nn\\
{\rm and}~~\g \dot{y}&=& \f{eB}{c} \dot{x} + \eta_y(t)\nn
\eea
where $\g$ is the friction coefficient due to the underlying bath and $\eta_x(t)$ and $\eta_y(t)$ are the concomitant random (stochastic) forces generated by the bath. The electrical drift mobility is defined as $e/\g$ for electrons. We assume as usual the random forces to be Gaussian white noise with $\langle \eta_{\alpha,\beta}(t)\rangle=0$ and  $\langle \eta_\alpha(t)\eta_\beta(t') \rangle=\eta_0^2 ~\delta (t-t')~ \delta_{\alpha\beta}$~. Here $\langle \ldots \rangle$ denotes average over the stochastic ensemble at time $t$. Solving  the above two linearly coupled equations for the velocities $\dot{x}$ and $\dot{y}$, we obtain 
\bea
\dot{x}&=&\f{\beta}{1+\beta^2} \{-\f{\alpha}{\beta}+\f{\eta_x(t)}{\beta
\g}-\f{\eta_y(t)}{\g} \}\label{xdot}\\
\dot{y}&=&\f{\beta}{1+\beta^2}\{ -\alpha+\f{\eta_x(t)}{\g}+\f{\eta_y(t)}{\beta \g}\}\label{ydot}
\eea 
with $eB/\g c=\beta$ and $eE/\g=\alpha$. In the overdamped limit, the information of the electron's motion is captured in the
probability distribution $P(x,y,t)$ of the x- and y-positions at time $t$. We are
interested in finding the marginal probability distributions $(P(x,t),~P(y,t))$ of
the displacements $x$ and $y$ respectively: $P(x,t)=\int P(x,y,t) dy$ and
$P(y,t)=\int P(x,y,t) dx$. From the van Kampen lemma \cite{nkumar85}, we can  write down a continuity equation for the marginal density $\pi(x,t)$ evolving stochastically under Eq.(\ref{xdot}) with $P(x,t)=\langle \pi(x,t) \rangle$ as
\bea
\f{\partial \pi (x,t)}{\partial t}&=&-\f{\partial}{\partial x}(\dot{x}~\pi (x,t))\nn\\
&=&\f{\beta}{1+\beta^2}\big\{\f{\alpha}{\beta}\f{\partial \pi (x,t)}{\partial
x}-\f{1}{\beta \g}\f{\partial}{\partial x}(\eta_x(t)\pi
(x,t))\nn \\
&+&\f{1}{\g}\f{\partial}{\partial x}(\eta_y(t)\pi (x,t))\big\}
\label{pidist}
\eea
The Fokker-Planck equation for the marginal probability distribution
$P(x,t)$ is now obtained by the noise averaging of Eq.(\ref{pidist}). We apply Novikov theorem \cite{nkumar85} for the Gaussian noise and find    
\bea
\f{\partial P(x,t)}{\partial t}=\f{\alpha}{1+\beta^2}\f{\partial P(x,t)}{\partial
x}+\f{\eta_0^2}{2 \g^2 (1+\beta^2)} \f{\partial^2 P(x,t)}{\partial x^2}
\label{fokk1}
\eea
The last equation describes the diffusion of electrons with drift. Similarly,  the Fokker-Planck equation for the marginal probability distribution $P(y,t)$:
\bea
\f{\partial P(y,t)}{\partial t}=\f{\alpha \beta}{1+\beta^2}\f{\partial
P(y,t)}{\partial x}+\f{\eta_0^2}{2 \g^2 (1+\beta^2)} \f{\partial^2 P(y,t)}{\partial y^2}\label{fokk2}
\eea
It is easier to solve the differential Eqs.(\ref{fokk1},\ref{fokk2}) by the Fourier transform method with initial conditions $P(x,t=0)=\delta(x)$ and
$P(y,t=0)=\delta(y)$. It is given as
\bea
P(x,t)=\sqrt{\f{\g^2(1+\beta^2)}{2\pi \eta_0^2t}}~{\rm exp}\Big
\{-\f{\g^2(1+\beta^2)}{2 \eta_0^2t}\Big(x+\f{\alpha}{1+\beta^2}t\Big)^2\Big \}\nn 
\eea
The marginal probability density $P(x,t)$ is a Gaussian distribution with mean $\langle x \rangle=
-\alpha t/(1+\beta^2)$ and variance $(\langle x^2 \rangle-\langle x
\rangle^2)=\eta_0^2t/\g^2(1+\beta^2)$. Here $v_{xd}=-\alpha/(1+\beta^2)$ is the
drift velocity along the $x-$direction. It can also be derived from Eq.(\ref{xdot}) by taking ensemble average on both sides. The above solution, of course, reduces to ordinary diffusion in the limit $E=0$ and $B=0$ with the identification $\eta_0^2=2D \g^2$, where $D$ is the diffusion constant. We can now readily derive the usual barotropic-type work fluctuation relation in the direction of electric field:
\bea
\f{P(x,t)}{P(-x,t)}={\rm exp}\Big \{-\f{2e\g E}{\eta_0^2}x\Big \}={\rm exp}\Big
\{-\f{e E x}{K_BT}\Big \}\label{FT1}
\eea
with the Einstein relation, $D\g=k_BT$. The interpretation of the above relation is straightforward. Though the system is always far from thermodynamic equilibrium (as the electron continuously dissipates energy in the environment), it reaches a mechanical equilibrium asymptotically under the combined effect (forcing) of the electromagnetic fields and the viscous drag acting in opposition. After the system attains the steady state, if the electron is at a 
position $x_1$, say, at $t=0$, then $P(x,t)$ is the conditional probability of finding the
particle at position $x_2$ at later time $t$ with the displacement $x=x_2-x_1$. Now, we reverse the positions keeping time direction as before and start at $t=0$ from the position $x_2$, then
$P(-x,t)$ is the conditional probability of finding the particle at position $x_1$ at time $t$. The main feature of the relation (\ref{FT1}) is that right-hand side is time independent. The relation (\ref{FT1}) can also be derived from the
eqilibrium probability distribution (depicting microscopic reversibility) \cite{Astumian06}. Using now the normalisation condition of the probability distribution, we obtain another useful relation from Eq.(\ref{FT1}): 
\bea
\Big \langle{\rm exp}\Big \{\f{e E x}{K_BT}\Big \} \Big
\rangle&=&\int_{-\infty}^{\infty}{\rm exp}\Big \{\f{e E x}{K_BT}\Big
\}P(x,t)dx\nn \\
&=&\int_{-\infty}^{\infty}P(-x,t)dx=1 \label{gFT1}
\eea  
The last two relations (Eq.(\ref{FT1}),Eq.(\ref{gFT1})) are similar to the generalised fluctuation-dissipation theorems which were derived in a very different context in Ref.\cite{Bochkov81}. 

Next,  we investigate the rather novel and interesting transverse fluctuations. Again, using the Fourier
transform method, we solve Eq.(\ref{fokk2}) for the marginal probability distribution $(P(y,t))$ of $y$ giving
\bea
P(y,t)=\sqrt{\f{\g^2(1+\beta^2)}{2\pi \eta_0^2t}}~{\rm exp}\Big
\{-\f{\g^2(1+\beta^2)}{2 \eta_0^2t}\Big(y+\f{\alpha \beta}{1+\beta^2}t\Big)^2\Big\}\nn 
\eea 
where the drift velocity along the $y-$direction is $v_{yd}=-\alpha \beta/(1+\beta^2)$.
The ratio of the probability densities for the transverse fluctuations is given
by
\bea
\f{P(y,t)}{P(-y,t)}={\rm exp}\Big \{-\f{e^2E By}{c\g K_BT}\Big \}={\rm exp}\Big \{-\f{eE y}{K_BT}\beta\Big \} \label{FT2}
\eea
where the interpretation of $P(y,t)$ and $P(-y,t)$ are similar to that for the x-direction.
We call the last relation ``Hall fluctuation'' theorem --it involves the product of the crossed electric and magnetic fields. The magnetic field, of course, does no work on a charge moving in the $xy$-plane. This cross ``Hall fluctuation'' effect can be related ultimately to the time-reversal symmetry breaking effect of the applied magnetic field, which is  different from the time-irreversibility introduced through the dissipation ($\g$). We can interpret it variously as arriving from the transverse Hall voltage (the Lorentz force), or effectively as a  magnetoresistance in the Hall geometry which occurs due to the enhancement of the path length of electron's motion caused by the magnetic field.  Finally, we give a generalised fluctuation-dissipation theorem-like expression in the presence of the magnetic field:
\bea
\Big \langle{\rm exp}\Big \{\f{eE y}{K_BT}\beta\Big \} \Big \rangle=1 \nn
\label{gFT2}
\eea
\begin{figure}
\begin{center}
\includegraphics[width=8.5cm]{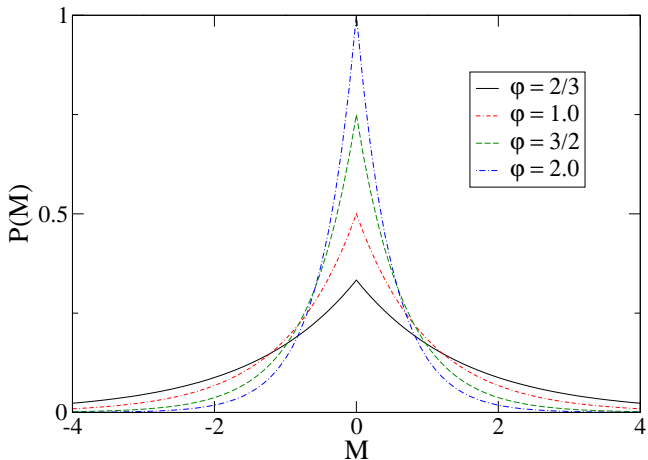}
\end{center}
\caption{(color online). Plot of the probability density $P(M)$ of Eq.(\ref{magdist}) vs. magnetic moment $M$ for different values of $\varphi$. Here $M$ is scaled by the Bohr magneton $(\mu_B)$.}
\label{mag1}
\end{figure}

$Diamagnetic~ fluctuations:$
Next, we turn to the diamagnetic fluctuations. We derive and briefly discuss the classical equilibrium fluctuations of the orbital diamagnetic moment (which is known to vanish identically in the mean). More specifically, we find the equilibrium probability density for the orbital magnetic moment of a charged particle in two dimensions for a normal magnetic field and a harmonic confinement. The probability density, when properly scaled, turns out to be universal and peaked about a zero mean value. Here, however,  we must retain the inertial effect (non-zero electron mass).

\begin{figure}
\begin{center}
\includegraphics[width=8.5cm]{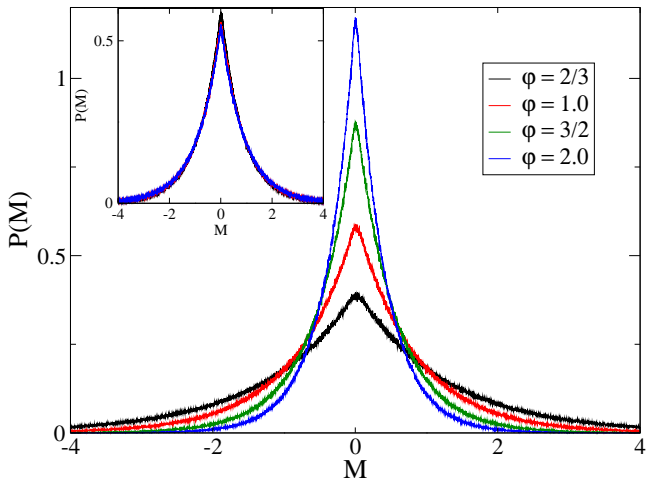}
\end{center}
\caption{(color online). Plot of the probability density $P(M)$ vs. magnetic moment $M$ obtained in simulation for same values of $\varphi$ as in FIG. \ref{mag1}. Peak of the curves decreases with decreasing $\varphi$. We have used $\g=1$, $m=1$ and again $M$ is scaled by $\mu_B$. Inset $P(M)$ for $\varphi=1.0$; and for different values of $\g$ and $B$ showing no effect.}
\label{mag2}
\end{figure}

Consider the orbital motion of the electron in the $xy$-plane for the normal external uniform magnetic field $B$ along $z-$directions and a harmonic confinement in the $xy$-plane of strength $k_0$. The Hamiltonian 
\bea
\mathcal{H}=\frac{1}{2m}(p_x-\frac{eBy}{2c})^2+\frac{1}{2m}(p_y+\frac{eBx}{2c})^2+\frac{1}{2}k_0(x^2+y^2) \nn
\eea
where we have used the symmetric Landau gauge $({\bf A}={\bf B} \times {\bf r}/2)$ for the vector potential ${\bf A}$. The orbital diamagnetic momemt can be expressed as
\bea
M(t)&=&\frac{-e}{2c}(x\dot{y}-y\dot{x})\nonumber \\
&=&\frac{-e}{2mc}\Big[x(p_y+\frac{eBx}{2c})-y(p_x-\frac{eBy}{2c})\Big] \label{magmom} 
\eea
We are interested in the asymptotic $(t \to \infty)$ distribution of $M(t)$. Now, we can evaluate the probability density $P(M)$ of classical $M$ in the limit $t \to \infty$ (i.e. in equilibrium) through the usual method of finding the equilibrium distribution at finite temperature $T$:
\bea
P(M)&=&\frac{1}{Z}\int_{-\infty}^{\infty}..\int_{-\infty}^{\infty}dx~ dy~ dp_x~ dp_y~e^{- \mathcal{H}/k_B T}\nn \\
&& \delta \Big(M+\f{e}{2mc}\Big[x(p_y+\frac{eBx}{2c})-y(p_x-\frac{eBy}{2c})\Big]\Big)\nn
\eea
Here $Z$ is a normalisation constant to be determined through $\int_{-\infty}^{\infty} P(M)dM=1$ (infact $Z$ is the equilibrium partition function for this model). After some simple algebra we obtain the probability density
\bea
P(M)=\frac{1}{2\mu_B}(\frac{\hbar \omega_0}{k_B T})~{\rm exp}\Big(-\frac{\hbar \omega_0}{k_B T}\f{|M|}{\mu_B}\Big) \label{magdist}
\eea
where $\mu_B=e\hbar/2mc$ is the Bohr magneton and $\omega_0=\sqrt{k_0/m}$. As expected $\la M \ra=0$, i.e., orbital diamagnetism is identically zero in a confined system. But, we do find finite thermal fluctuations of the orbital diamagnetic moment in confinement. In FIG. \ref{mag1}, we plot $P(M)$ for different values of the dimentionless parameter $\varphi=\hbar \omega_0/k_BT$. It is interesting to note that the classical partition function $(Z)$, and hence the free energy $(-k_BT {\rm ln}Z)$ are independent of the magnetic field $B$ giving vanishing orbital diamagnetism in the mean. But the moments of the orbital magnetic fluctuation of even order are all non-zero. Clearly, then these orbital magnetic moments can not be derived through the usual field derivatives of the classical free energy as would be the case for ``permanent magnetic moments'' intrinsic to the particles. The classical equilibrium simulation of the above model with the Langevin heat bath can be shown to be consistent with the above thermal fluctuations of the magnetic moments. We simulate the coupled set of Langevin equations 
\bea
m\ddot{x}&=&-\g \dot{x}-\f{eB}{c} \dot{y}-k_0x +\eta_x(t) \nn\\
{\rm and}~~m\ddot{y}&=&-\g \dot{y}+\f{eB}{c} \dot{x}-k_0y + \eta_y(t)\nn
\eea   
where again $\g$ and $\eta$ are respectively the friction and the noise, related through $\langle \eta_\alpha(t)\eta_\beta(t') \rangle= 2\g k_B T\delta (t-t')~ \delta_{\alpha\beta}$. Here we use the velocity-Verlet algorithm for the time evolution of the above equations and find steady-state equilibrium distribution of the magnetic moment given by Eq.(\ref{magmom}). We plot the distribution $P(M)$ in FIG. \ref{mag2} for the same parameter values of $\varphi$ as in FIG. \ref{mag1}. Also, we confirm through our simulation that the $P(M)$ is independent of $\g$ and $B$ (see inset of FIG. \ref{mag2}).

$Concluding~ remarks:$
In this paper we have re-derived a barotropic-type work-fluctuation relation along with a new  transverse fluctuation relation for the case of the classical motion of a charged particle in static homogeneous crossed magnetic and electric fields in the presence of dissipation. This is interesting inasmuch as  classically the magnetic field does no work on a moving charge particle. Recently, there have been some studies of fluctuation theorems \cite{amj07} in time-varying electromagnetic fields. But, our approach and the context are quite different from these. As our treatment is based on the classical Langevin equations involving stochastic fluctuating forces  and the concomitant dissipation that neglect quantum statistics and the band structure effects \cite{eric06}, it is expected to be appropriate for a material system which is electronically non-degenerate, i.e., has a low carrier density at a relatively high temperature, and has a low carrier mobility. Clearly, a polar semiconductor with strong electron-phonon interaction is indicated. The inertial mass occurring in our Langevin equations is, of course, to be replaced by the effective mass relevant to the bottom (top) of the conduction (valence) band for the electron (hole) as the band is expected to be close to being parabolic there. We have also derived the probability density for the classical diamagnetic moment giving non-zero orbital magnetic moments of even order. The latter can not be obtained as field derivatives of the equilibrium free energy which is clasically field independent.

One of us (D.R.) thanks A. Dhar and R. Marathe for fruitful discussions.

\end{document}